\documentclass[reprint,aps,prl,twocolumn,superscriptaddress]{revtex4-1}
\usepackage[dvips]{graphicx}
\usepackage{color}
\usepackage{CJK}

\begin{document}
\begin{CJK*}{UTF8}{mj}

\title{Novel Crossover in Coupled Spin Ladders}

\affiliation{Laboratoire National des Champs Magn\'etique Intenses, LNCMI-CNRS (UPR3228), EMFL, \\ UGA, UPS, and INSA, Bo\^{i}te Postale 166, 38042, Grenoble Cedex 9, France}
\affiliation{Neutron Scattering and Magnetism, Laboratory for Solid State Physics, ETH, Z\"urich, Switzerland}
\affiliation{Laboratory for Quantum Magnetism, Institute of Physics, Ecole Polytechnique F\'{e}derale de Lausanne (EPFL), CH-1015 Lausanne, Switzerland}

\author{M. Jeong (정민기)}
\email{minki.jeong@gmail.com}
\affiliation{Laboratoire National des Champs Magn\'etique Intenses, LNCMI-CNRS (UPR3228), EMFL, \\ UGA, UPS, and INSA, Bo\^{i}te Postale 166, 38042, Grenoble Cedex 9, France}
\affiliation{Laboratory for Quantum Magnetism, Institute of Physics, Ecole Polytechnique F\'{e}derale de Lausanne (EPFL), CH-1015 Lausanne, Switzerland}
\author{H. Mayaffre}
\affiliation{Laboratoire National des Champs Magn\'etique Intenses, LNCMI-CNRS (UPR3228), EMFL, \\ UGA, UPS, and INSA, Bo\^{i}te Postale 166, 38042, Grenoble Cedex 9, France}
\author{C. Berthier}
\affiliation{Laboratoire National des Champs Magn\'etique Intenses, LNCMI-CNRS (UPR3228), EMFL, \\ UGA, UPS, and INSA, Bo\^{i}te Postale 166, 38042, Grenoble Cedex 9, France}
\author{D. Schmidiger}
\affiliation{Neutron Scattering and Magnetism, Laboratory for Solid State Physics, ETH, Z\"urich, Switzerland}
\author{A. Zheludev}
\affiliation{Neutron Scattering and Magnetism, Laboratory for Solid State Physics, ETH, Z\"urich, Switzerland}
\author{M. Horvati\'c}
\email{mladen.horvatic@lncmi.cnrs.fr}
\affiliation{Laboratoire National des Champs Magn\'etique Intenses, LNCMI-CNRS (UPR3228), EMFL, \\ UGA, UPS, and INSA, Bo\^{i}te Postale 166, 38042, Grenoble Cedex 9, France}

\begin{abstract}
We report a novel crossover behavior in the long-range-ordered phase of a prototypical spin-$1/2$ Heisenberg antiferromagnetic ladder compound $\mathrm{(C_7H_{10}N)_2CuBr_4}$. The staggered order was previously evidenced from a continuous and symmetric splitting of $^{14}$N NMR spectral lines on lowering temperature below $T_c\simeq 330$ mK, with a saturation towards $\simeq 150$ mK. Unexpectedly, the split lines begin to further separate away below $T^*\sim 100$ mK while the line width and shape remain completely invariable. This crossover behavior is further corroborated by the NMR relaxation rate $T_1^{-1}$ measurements. A very strong suppression reflecting the ordering, $T_1^{-1}\sim T^{5.5}$, observed above $T^*$, is replaced by $T_1^{-1}\sim T$ below $T^*$. These original NMR features are indicative of unconventional nature of the crossover, which may arise from a unique arrangement of the ladders into a spatially anisotropic and frustrated coupling network.
\end{abstract}

\date{\today}
\maketitle
\end{CJK*}

Spin ladders in a magnetic field are a paradigmatic model in quantum magnetism and many-body physics \cite{Giamarchi99PRB, Giamarchi}. For instance, a spin-$1/2$ Heisenberg antiferromagnetic (AFM) ladder in a field between the two critical values, $H_{c1}$ and $H_{c2}$, hosts as the ground state a Tomonaga-Luttinger liquid (TLL), a state universal to interacting quantum particles in one dimension (1D) with gapless excitations \cite{Giamarchi, Haldane81JPC, Haldane81PRL}. When the ladders are embedded in real material, a weak residual coupling between them comes into play at sufficiently low temperatures, and this dimensional crossover manifests itself as a second-order phase transition into a canted $XY$ ordered phase \cite{Giamarchi99PRB}. This 3D long-range-ordered (LRO) phase is described as a Bose-Einstein condensate (BEC) of magnons \cite{Giamarchi99PRB, Nikuni00PRL, Giamarchi, Giamarchi08NPhys, Bouillot11PRB}. The transition between the two canonical quantum phases, 1D TLL and 3D BEC, has been successfully demonstrated with a metal-organic spin-ladder compound $\mathrm{(C_5H_{12}N)_2CuBr_4}$, known as BPCB, by NMR \cite{Klanjsek08PRL} and neutron diffraction \cite{Thielemann09PRB}. The same class of transition has been observed since then in an increasing number of quasi-1D spin systems of magnetic insulators including bond-alternating AFM chains \cite{Willenberg15PRB}, and also in ultracold atoms trapped in an array \cite{Vogler14PRL}.

\begin{figure}
    \centering
    \includegraphics[width=0.5\textwidth]{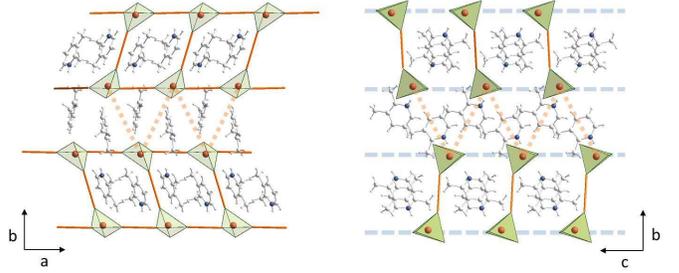}
    \caption{Crystal structure of DIMPY where orange balls in green tetrahedra represent spin-$1/2$ Cu ions of $\mathrm{CuBr}_4$ units and solid lines represent predominant exchange pathways forming a ladder-like network. Broken lines represent much weaker couplings between the ladders. (a) View presenting the ladders side by side. (b) View along the ladder direction showing the coupling of the ladders along the $b$ and $c$ axes.}\label{exchange}
\end{figure}

\begin{figure*}
    \centering
    \includegraphics[width=1\textwidth]{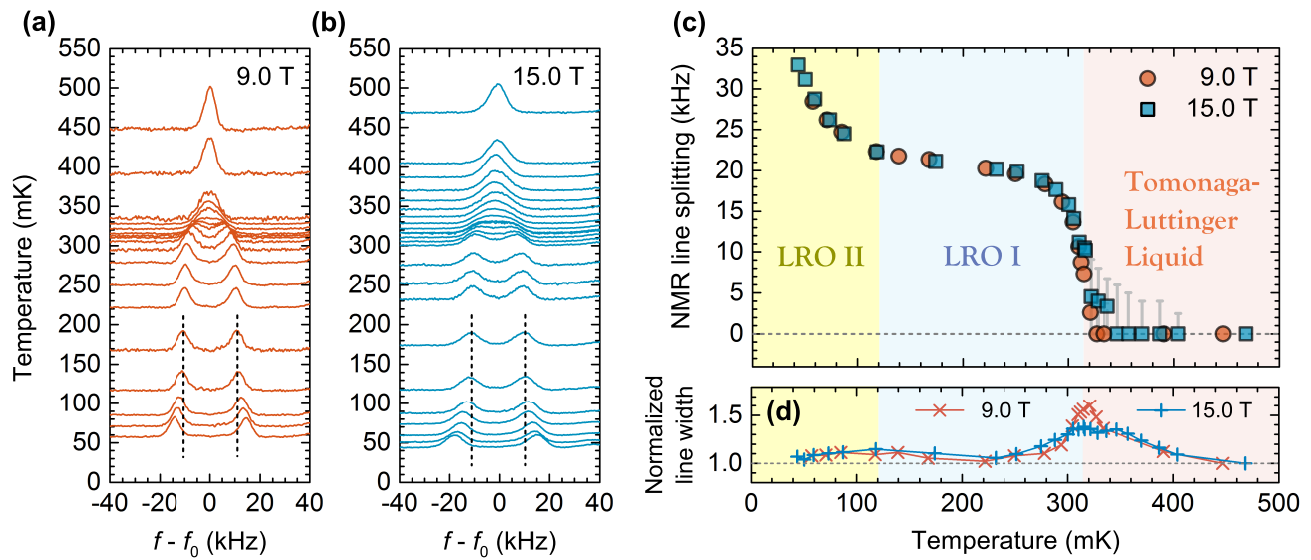}
    \caption{(a) $^{14}$N NMR spectra as a function of temperature in an applied field of 9.0 T and (b) in 15.0 T. Dashed vertical lines correspond to the first saturation of the line splittings at around 150 mK on cooling. (c) The line splitting as a function of temperature, where three different regimes are identified and presented using different background colors. (d) Normalized line width as a function of temperature.}\label{splitting}
\end{figure*}

Recently, another metal-organic spin-ladder compound $\mathrm{(C_7H_{10}N)_2CuBr_4}$, known as DIMPY, has attracted much attention \cite{Shapiro07JACS, White10PRB, Hong10PRL, Schmidiger11PRB, Ninios12PRL, Schmidiger12PRL} as a unique example for a strong-leg regime, i.e., $J_\mathrm{leg}/J_\mathrm{rung}=1.7$ where $J_\mathrm{leg}=16.5$ K (see Fig.~\ref{exchange} for crystal structure) \cite{Shapiro07JACS, White10PRB, Hong10PRL, Schmidiger11PRB, Ninios12PRL, Schmidiger12PRL} with experimentally accessible $H_{c1}\simeq 2.5$ T and $H_{c2}\simeq 29$ T. The single-ladder (1D) Hamiltonian of DIMPY has been thoroughly determined by using inelastic neutron scattering in conjunction with the density matrix renormalization group (DMRG) calculations and bulk measurements \cite{White10PRB, Hong10PRL, Schmidiger11PRB, Ninios12PRL, Schmidiger12PRL}. The low-energy excitations in a single-ladder limit, probed via inelastic neutron scattering \cite{Ninios12PRL, Schmidiger12PRL, Schmidiger13PRL, Povarov15PRB} and NMR relaxation \cite{Jeong13PRL, Jeong16PRL}, were shown to agree with the TLL predictions \cite{Giamarchi99PRB, Giamarchi}. Moreover, specific-heat anomalies \cite{Ninios12PRL, Schmidiger12PRL} observed typically around $T_c\sim 300$ mK in a magnetic field $H>H_{c1}$ were shown to correspond to an onset of a staggered LRO due to weak interladder couplings \cite{Jeong13PRL}. Therefore, DIMPY, together with a weak-leg ladder representative compound BPCB ($J_\mathrm{leg}/J_\mathrm{rung}=0.28$) \cite{Bouillot11PRB, Klanjsek08PRL, Ruegg08PRL}, is considered to provide a complete experimental toolkit for exploring the physics of coupled spin ladders in a field \cite{Bouillot11Geneve, Schmidiger13PRB, Schmidiger14ETH, Jeong16PRL, Steinigeweg16PRL}.

We report in this Letter a new set of NMR observation on DIMPY which defies the standard paradigm \cite{Giamarchi99PRB} of the coupled spin ladders in a field. We discover an unexpected crossover taking place around $T^*\sim 100$ mK, where upon cooling the size of the seemingly saturated ordered moments grows again and the low-energy excitations change the nature. We present the original NMR signatures of the crossover and discuss a possible origin in light of the recent theory~\cite{Furuya16PRB}.

A single-crystal sample was directly put into a $^3$He-$^4$He mixture of a dilution refrigerator to ensure a good thermal contact. $^{14}$N (nuclear spin value $I=1$) NMR experiments were performed using a standard pulsed spin-echo technique. The spectrum was obtained by performing a Fourier transform of the spin echo signal that follows an excitation and refocusing NMR pulses. NMR spin-lattice relaxation rate, $T_1^{-1}$, was obtained by a saturation-recovery method, using the theoretical relaxation function for $I=1$ nuclei, $M(t)/M_0=1-0.25\exp (-(t/T_1)^\alpha) - 0.75\exp (-(3t/T_1)^\alpha)$, where $M(t)$ is nuclear magnetization, $t$ is a time interval between the saturation pulse and the echo pulses, and $M_0$ the nuclear magnetization in equilibrium ($t\rightarrow \infty$). The stretch exponent $\alpha$ was introduced to describe distribution of $T_1^{-1}$ values. The saturation of nuclear magnetization was achieved by using a single pulse as long as $10\sim 20\,\mu s$ to reduce the excitation power so that unwanted heating effects were avoided.

Figure \ref{splitting}(a) and (b) show the $^{14}$N NMR line shape as a function of temperature in an applied field of 9.0 and 15.0 T, respectively. In both fields, a spectral line at high temperatures becomes broadened as temperature is lowered, and then splits symmetrically into two lines across $T_c\simeq 330$ mK \cite{Jeong13PRL}. This splitting reflects the growth of the staggered transverse ($\perp H$) moments, i.e., the order parameter (OP). Figure~\ref{splitting}(c) plots temperature evolution of the splitting which tends to saturate as temperature approaches 150 mK. However, as temperature is further lowered across $T^*\sim 100$ mK, the split lines begin to separate further away symmetrically. At the lowest measured temperature of 40 mK the splitting becomes 33 kHz, which is 50 \% larger than the 22 kHz observed at $\sim 150$ mK.

The NMR lines have a Gaussian shape over the measured temperature and field ranges, except close to $T_c$ where the line shape can be decomposed into two superimposed Gaussians (Fig.~\ref{splitting}(a) and (b)). The line widths at high temperatures above 400 mK are 4.4 and 6.9 kHz in 9 and 15 T, respectively, meaning that the line broadening scales with the field and is thus of magnetic origin. When temperature is lowered across $T_c$, the line broadens on top of the splitting which is a hallmark of magnetic ordering transition. On the other hand, the line shape and width remain almost completely intact across $T^*$. Figure \ref{splitting}(d) plots the line width normalized by the high temperature value as a function of temperature. The overall spectral features are practically indistinguishable between 9 and 15 T.

\begin{figure*}
    \centering
    \includegraphics[width=0.8\textwidth]{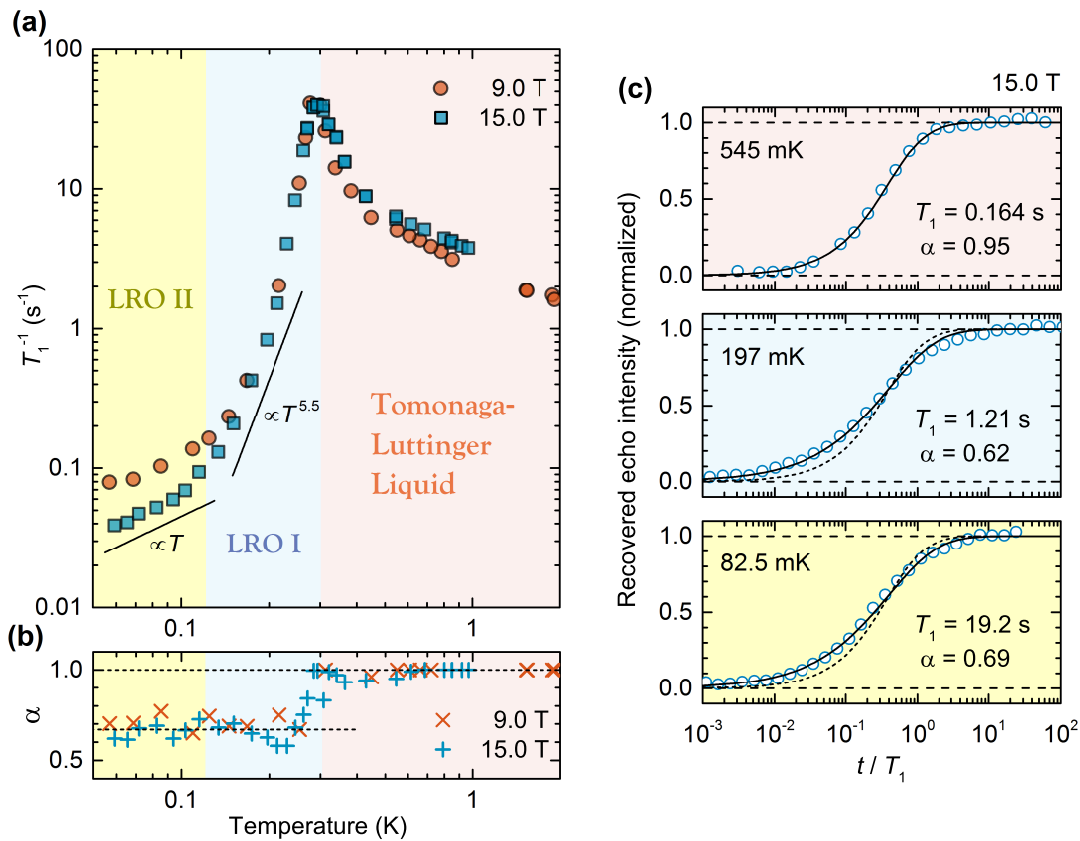}
    \caption{(a) $^{14}$N NMR relaxation rate $T_1^{-1}$ as a function of temperature in 9.0 T and 15.0 T. (b) Temperature evolution of the stretch exponent $\alpha$ in the relaxation function. (c) Nuclear magnetization recovery curves for 15 T at 545 mK, 197 mK, and 82.5 mK, from upper to lower panels. Solid lines are fits using the theoretical relaxation function with varying $\alpha$ while dotted lines are with $\alpha=1$ for comparison.}\label{T1ex}
\end{figure*}

The crossover behavior in the spectrum across $T^*$ is further corroborated by the relaxation rate measurements. Figure~\ref{T1ex}(a) shows $T_1^{-1}$ as a function of temperature in 9.0 T and 15.0 T. Note that $T_1^{-1}$ probes Cu$^{2+}$ electron spin fluctuations in the low energy limit. At high temperatures in the TLL regime, $T_1^{-1}$ increases with lowering temperature by 1D quantum-critical fluctuations \cite{Jeong13PRL}. As temperature further approaches $T_c$, the $T_1^{-1}$ increases even more rapidly by the addition of thermal-critical fluctuations, which is another hallmark of magnetic ordering transition. Then, a very strong suppression of $T_1^{-1}$, by more than two orders of magnitude, follows the peak at $T_c$ as temperature is lowered below 300 mK. In the temperature range where the OP is apparently saturated, we find $T_1^{-1}\sim T^{5.5}$. A similar suppression has been observed in other quasi-low-dimensional quantum magnets below the ordering transition \cite{Mayaffre00PRL}. However, as temperature is further lowered across $T^*$, the $T_1^{-1}$ begins to bend out from the strong suppression, and roughly follows $T_1^{-1}\propto T$ behavior. In addition, $T_1^{-1}$ in 9 T is roughly twice larger than the one in 15 T in this regime.

Figure~\ref{T1ex}(b) plots the stretch exponent $\alpha$ used to fit the nuclear magnetization recovery curves shown in Fig.~\ref{T1ex}(c) by the theoretical relaxation function. This exponent is indicative of local magnetic inhomogeneity. For instance, $\alpha$ remains practically 1 above $T_c$, indicating a homogeneous magnetic environment. When temperature is lowered below $T_c$, $\alpha$ drops down to $0.6\sim 0.7$, which indicates the development of local dynamic inhomogeneity or distribution of $T_1^{-1}$ values. Although the temperature dependence of $T_1^{-1}$ changes across $T^*$, this is not accompanied by any noticeable modification of $\alpha$.

Let us now discuss the consequences of the observed NMR signatures of the crossover between the two low temperature regimes, above and below $\sim$100 mK, which we label as LRO I and LRO II, respectively (see Fig.~\ref{splitting}(c) and \ref{T1ex}(a)). Since the width of the NMR lines does not increase nor $T_1^{-1}$ shows a peak across $T^*$, one can rule out a symmetry-breaking, continuous phase transition accompanied by critical fluctuations. In addition, quadrupolar splitting (not shown), which probes crystalline electric field gradient, does not change over the whole temperature range, pointing to the absence of any structural change. Indeed, at such low temperatures phonon modes are likely to be completely quenched. We thus simply associate the increasing NMR line separation below $T^*$ with additional size growth of the OP. The enhanced $T_1^{-1}$ at high temperatures in the TLL and thermal-critical regimes are suppressed below $T_c$ as the fluctuations associated with the magnetic moments are suppressed and the long range order develops.  Regarding the emergence of the $T_1^{-1}\propto T$ behavior below $T^*$, there are two possible cases: one is that the fluctuation spectrum itself changes qualitatively, while the other is that the temperature-linear behavior is intrinsic to the ordered phases but only revealed once the enhanced fluctuations are quenched out.

We note that our NMR observations have certain correspondence to the recent theory on spatially anisotropic Heisenberg antiferromagnets, with DIMPY as a specific example \cite{Furuya16PRB}. The theory bases the argument on a uniquely anisotropic interladder coupling network of DIMPY. Naively, from the distances between the magnetic $\mathrm{Cu}^{2+}$ ions, one could expect that the interladder exchange interactions would be stronger along the $c$ direction than the $b$ direction (Fig.~\ref{exchange}). Moreover, presuming the interactions are antiferromagnetic along the $b$ axis, they would be frustrated such that the effective coupling strength is even further reduced. Thus, depending on the energy scale defined by the temperature, the magnetic lattice of DIMPY may be considered as a 1D ladder, a 2D net of coupled ladders, and a 3D stack. Taking into account this hierarchical coupling strengths, the analytical and numerical results predict on cooling a transition from the TLL into a quasi-2D regime of the ordered phase, which in fact does present a full 3D coherence, but has reduced OP because of strong fluctuations. This is followed by a crossover to a ``true'' 3D regime, where these fluctuations are frozen and the OP is thus bigger. In this crossover, the OP monotonically increases on cooling \cite{Furuya16PRB}, which apparently corresponds to our experimentally observed NMR line splitting in the LRO II regime. However, the theoretically predicted temperature dependence of the OP does not really present a plateau for the quasi-2D regime, and is thus somewhat different from the one observed in the LRO I regime.

On the other hand, the NMR line shape is in sharp contrast to what is expected from the above theory. The main characteristic of the predicted quasi-2D ordered regime is that the phase (orientation) of the local OP value is only very weakly correlated between the planes, which connects to a Berezinskii-Kosterlitz-Thouless phase~\cite{Berezinskii1SJETP71, Kosterlitz73JPC} in the limit of an isolated plane. In the above theory  \cite{Furuya16PRB}, presuming a clean system without impurities, very weak interplane coupling and thus the strong phase fluctuations give rise to a small OP in the quasi-2D regime. Real compounds always have impurities which act as the pinning centers, so that the \emph{local} OP phase is expected to be pinned in different directions throughout the sample.  As the NMR hyperfine coupling is strongly sensitive to the orientation of the OP, different local orientations correspond to different NMR line positions. Thus, in the quasi-2D regime having relatively uncorrelated planes, one expects a broadly distributed NMR spectrum. In the 3D regime at lower temperature, the OP is growing because the 3D coupling suppresses the strong phase fluctuation. In real compounds, this corresponds to the local OP phase becoming homogeneously locked to the same value throughout the sample, meaning that the NMR line width should shrink to its normal value (as above $T_c$). In contrast to these prediction, the NMR lines in DIMPY do not broaden nor change their shape upon entering the ordered phases, and we therefore conclude that the OP has the same average orientation throughout the whole sample, in both LRO I and LRO II regimes.

The absence of the temperature dependence of the NMR line widths suggests an alternative scenario: some (small) anisotropy defining a preferential direction in real materials, particularly in crystals of low symmetry as DIMPY, may play a role. This would fix the OP phase and thus ensure a full 3D coherence at all temperature, leading to a temperature-independent NMR line width. Furthermore, depending whether the temperature is higher or lower than the energy scale of this anisotropy, the OP phase fluctuations will be either strong or frozen to the optimal direction; the corresponding average OP value will be thus reduced at higher temperature and will grow towards its full size at lower temperature, as observed by the NMR line splitting. Moreover, there is an obvious candidate for the anisotropy in DIMPY: the local crystal symmetry allows uniform Dzyaloshinkii-Moriya interactions, whose coupling strength is estimated to be as large as 310~mK, of the same size as the $T_c$ value \cite{Ozerov15PRB, Glazkov15PRB}. Inclusion of such anisotropic terms into the theory is challenging, and beyond the scope of this work.

Unfortunately, the anisotropy scenario may have difficulty in explaining the low temperature $T_1^{-1}$ data. Across $T^*$, the accompanying freezing of spin fluctuations would lead to the quenching of $T_1^{-1}$ relaxation rate. However, $T_1^{-1}$ is not quenched at low temperature, but is rather maintained or enhanced. One possible explanation is that the low temperature relaxation is due to \emph{accidentally} concomitant setting in of the impurity relaxation. This may indeed be field dependent as observed, as stronger field reduces impurity spin fluctuations. Another possibility is that the crossover comprises ``quasi-critical'' enhanced fluctuations which could give rise to the observed $T_1^{-1}$ behavior (though the field dependence would be subject to the actual model for the crossover). On the other hand, it is also interesting to note that the $T_1^{-1}\propto T$ behavior found in the LRO II is precisely the one that has been, on general grounds, theoretically predicted for a 3D BEC phase of weakly-coupled spin-$1/2$ Heisenberg AFM ladders \cite{Giamarchi99PRB}.


To conclude, the magnetized spin-ladder compound DIMPY displays a novel type of crossover with temperature in the LRO phase. We have shown that on cooling across the crossover the seemingly saturated order parameter amplitude grows again and, moreover, the low-energy excitations become strongly modified so that NMR relaxation rate becomes linear in temperature. These original observations have certain correspondence to the recent theory on spatially anisotropic Heisenberg antiferromagnets showing the crossover between the quasi-2D and the 3D ordered phases \cite{Furuya16PRB}. Admittedly, the correspondence is not complete, which calls for future theoretical and experimental studies. We hope that our new finding will help to elucidate the intriguing manifestation of effective dimensionality and frustration, and their interplay.

\begin{acknowledgments}
We acknowledge fruitful discussions with T. Giamarchi, S. C. Furuya and N. Laflorencie. This work was supported by the French ANR project BOLODISS (Grant No. ANR-14-CE32-0018). M.J. is grateful for the support of European Commission through Marie Sk{\l}odowska-Curie Action COFUND (EPFL Fellows), and Swiss National Science Foundation through Korean-Swiss Science and Technology Programme.
\end{acknowledgments}     

\bibliography{bib_QSpinSystems}

\end{document}